\documentclass[reprint,amsmath,amssymb,aps,]{revtex4-2}

\usepackage{graphicx}
\usepackage{dcolumn}
\usepackage{bm}
\usepackage[utf8]{inputenc}

\usepackage{lineno}
\usepackage{xcolor}
\usepackage{siunitx}
\usepackage{pgfplotstable} 
\pgfplotsset{compat=1.18}
\usepackage{booktabs} 
\usepackage{multirow}
\usepackage{tabularx}
\usepackage[normalem]{ulem}
\usepackage{makecell}
\usepackage{appendix}
\usepackage{hyperref}

\newcommand{\hyy}{$H \rightarrow \gamma\gamma$}

\begin{document}
\preprint{APS/123-QED}

\title{Pretrained Event Classification Model for High Energy Physics Analysis}

\thanks{{Code: \href{https://huggingface.co/HWresearch/GNN4Colliders}
{\nolinkurl{https://huggingface.co/HWresearch/GNN4Colliders}}.}}

\author{Joshua Ho}
\affiliation{
    Department of Physics, University of California, Berkeley, CA 94720\\
    Physics Division, Lawrence Berkeley National Laboratory, Berkeley CA 94720
}

\author{Ryan Roberts}
\affiliation{
    Department of Physics, University of California, Berkeley, CA 94720\\
    Physics Division, Lawrence Berkeley National Laboratory, Berkeley CA 94720
}

\author{Shuo Han}
\affiliation{
    Department of Physics, University of California, Berkeley, CA 94720\\
    Physics Division, Lawrence Berkeley National Laboratory, Berkeley CA 94720
}

\author{Haichen Wang}
\affiliation{
    Department of Physics, University of California, Berkeley, CA 94720\\
    Physics Division, Lawrence Berkeley National Laboratory, Berkeley CA 94720
}
\date{\today}

\begin{abstract}
We introduce a foundation model for event classification in high-energy physics, built on a Graph Neural Network architecture and trained on 120 million simulated proton-proton collision events spanning 12 distinct physics processes. The model is pretrained to learn a general and robust representation of collision data using challenging multiclass and multilabel classification tasks. Its performance is evaluated across seven event classification tasks, which include new physics processes not encountered during pretraining as well as ATLAS Open Data to demonstrate generalizability across different simulation frameworks, from Delphes fast simulation to full ATLAS detector simulation. Fine-tuning the pretrained model significantly improves classification performance, particularly in scenarios with limited training data, demonstrating gains in both accuracy and computational efficiency. To investigate the underlying mechanisms behind these performance improvements, we employ a representational similarity evaluation framework based on Centered Kernel Alignment. This analysis reveals that encoder-stage representations of the fine-tuned model remain similar to those of the baseline, while intermediate graph processing layers diverge substantially, indicating that fine-tuning preserves general-purpose encoders while developing fundamentally different message-passing pathways to arrive at superior task performance.

\end{abstract}

\maketitle

\section{Introduction}
Machine learning has become a ubiquitous tool in particle physics, employed in a variety of tasks including triggering, simulation, reconstruction, and offline analysis.  While its utility spans classification, regression, and generative tasks, the current paradigm of developing machine learning models from scratch for each specific application presents several challenges. This approach not only demands specialized expertise and substantial computing resources but can also result in suboptimal performance due to limited training data. The from-scratch development of models necessitates individual validation studies to ensure that neural networks utilize well-modeled information from training samples, whether derived from Monte Carlo simulations or control samples from experimental data.

Foundation models offer a promising direction to address these limitations. These models, pretrained on large, diverse datasets across various tasks, provide robust and general representations of underlying data structures. Notable examples in other fields include GPT-4~\cite{openai2024gpt4technicalreport} and BERT~\cite{DBLP:journals/corr/abs-1810-04805} in natural language processing, Stable Diffusion~\cite{DBLP:journals/corr/abs-2112-10752,podell2023sdxlimprovinglatentdiffusion} in image processing, and AlphaFold~\cite{jumper2021alphafold} in structural biology. The foundation model approach offers several advantages for particle physics applications: reduced computing resources for fine-tuning~\cite{DBLP:journals/corr/YosinskiCBL14} compared to training from scratch, superior performance on specific tasks (particularly with limited training data), and potentially simplified validation procedures as the pretrained representations have already been studied on a broad set of physics processes, providing a foundation that may inform but does not replace task-specific validation.

Current literature on pretrained models for particle physics can be categorized based on the data representation they handle. Models operating on particle- or event-level numerical data use features like particle four momenta or jets, leveraging self-supervised or generative methods to learn versatile representations. Detector-focused model operates on high-dimensional responses such as calorimeter deposits or pixel hits, employing geometry-aware techniques for accurate simulation and analysis. Finally, models using textual or code representations apply large language model architectures to integrate domain knowledge, enabling tasks like question answering and code generation. 

Recent studies have begun exploring foundation models tailored to particle physics data, which has a variety of distinct structures and properties across many experiments and data processing stages, including particle-level \& event-level numeric data ~\cite{wildridge2024bumblebeefoundationmodelparticle, katel2024learningsymmetryindependentjetrepresentations, golling2024maskedparticlemodelingsets, mikuni2024omnilearnmethodsimultaneouslyfacilitate, harris2024resimulationbasedselfsupervisedlearningpretraining, Birk_2024, vigl2024finetuningfoundationmodelsjoint}, 
detector-level \& geometry-aware data ~\cite{araz2024pointcloudbaseddiffusionmodels, Liu_2023, Hashemi_2024, huang2024languagemodelparticletracking}, and 
textual or code data ~\cite{zhang2024xiwubasisflexiblelearnable}.

While recent work has demonstrated the promise of foundation models in high-energy physics, existing approaches differ from ours in important ways. The majority of prior foundation models operate at the jet level, treating individual jets as the primary object of study. MPM \cite{leigh2024tokenizationneededmaskedparticle} and OmniJet-$\alpha$ \cite{Birk_2024} focus on learning representations of jet constituents using masked modeling and autoregressive generation respectively, while OmniLearn \cite{mikuni2024omnilearnmethodsimultaneouslyfacilitate} demonstrates strong cross-detector and cross-task generalization across a range of final states. RS3L \cite{harris2024resimulationbasedselfsupervisedlearningpretraining} proposes a self-supervised contrastive learning strategy at the jet level using re-simulation-based data augmentation. The work of Vigl et al. \cite{vigl2024finetuningfoundationmodelsjoint} demonstrates the value of pretraining and fine-tuning workflows for joint reconstruction and analysis optimization in HEP, but focuses on a specific search topology rather than developing a general-purpose event-level representation. Bumblebee \cite{wildridge2024bumblebeefoundationmodelparticle} is perhaps the closest in spirit to our approach, employing a BERT-inspired architecture at the event level by embedding particle four-vectors with permutation invariance. However, Bumblebee uses a Transformer-based architecture and a generative pretraining objective. In contrast, our model uses a Graph Neural Network architecture that explicitly encodes relational structure between particles in the event, and is pretrained on a substantially larger and more diverse dataset of 120 million events spanning 12 physics processes using discriminative multiclass and multilabel objectives. These distinctions make our approach complementary to existing jet-level foundation models, addressing the event-level classification setting that prior work has not systematically explored. We further evaluate generalizability across simulation frameworks, from Delphes fast simulation to full ATLAS detector simulation, a setting not addressed by prior work.

This paper presents a foundation model designed specifically for collider event-level data. In modern collider experiments, final-stage analysis processes information from reconstructed objects that either directly correspond to particles in collision final states (such as leptons and photons) or serve as proxies (such as jets and missing transverse energy). While traditional approaches often relied on ``high-level'' variables calculated from object features, recent trends favor direct input of event objects and their features into neural networks for analysis tasks. A notable example is Ref.~\cite{ATLAS:2023ajo}, which established the observation of simultaneous production of four top quarks with the ATLAS experiment by employing a graph neural network (GNN) architecture to process event-level object information.

We present foundation models that adopt an architecture similar to that used in Ref.~\cite{ATLAS:2023ajo}. Our models are pretrained using either multiclass classification or multilabel learning tasks across 12 distinct physics processes and 120 million simulated proton-proton collision events. We evaluate these models through fine-tuning and testing on seven classification tasks, including novel processes not seen during pretraining as well as ATLAS Open Data, the latter serving to demonstrate generalizability across simulation frameworks. Our analysis benchmarks the models' performance improvements, their scaling behavior with training sample size, and computational efficiency. To investigate the mechanisms underlying these improvements, we employ a representational similarity framework based on Centered Kernel Alignment, revealing that pretrained and scratch-trained models develop similar intermediate representations while diverging at the task-specific output level. This work represents the first prototype of a foundation model operating on collider final-state object data.

We use the term ``foundation model'' in the sense of a model trained on broad and diverse data at scale and adapted to a wide range of downstream tasks, without requiring very large model capacity or scaling-driven emergent behavior. In the High Energy Physics (HEP) analysis ecosystem, where task-specific classifiers are typically trained from scratch on limited simulated samples, a model pretrained on 120 million events spanning 12 physics processes and successfully adapted to seven downstream tasks across multiple simulation frameworks represents a qualitatively different paradigm, and we argue this paradigm is appropriately characterized as a foundation model approach in this context.

\section{Data Samples}
\label{sec:data}

To provide a diverse set of physics processes for the pretraining, we use Madgraph@NLO 2.7.3~\cite{Alwall:2014hca} to generate proton-proton collision events at next-to-leading order (NLO) in Quantum Chromodynamics (QCD). We generate 12 distinct Standard Model (SM) physics processes, including six major Higgs boson production mechanisms: gluon fusion production ($ggF$), vector boson fusion ($VBF$), associated production of the Higgs boson with a W boson ($WH$) or a Z boson ($ZH$), associated production of the Higgs boson with a top-quark pair ($t\bar{t}H$), and associated production of the Higgs boson with a single top quark and a forward quark ($tHq$). Additionally, we simulate six top quark production processes: single top production, top-quark pair production ($t\bar{t}$), top quark pair production in association with a pair of photons ($t\bar{t}\gamma\gamma$), associated production of a top-quark pair with a W boson ($t\bar{t}W$), simultaneous production of three top quarks ($t\bar{t}t$), and simultaneous production of four top quarks ($t\bar{t}t\bar{t}$). In these samples, the Higgs boson and top quarks decay inclusively. These 12 Higgs and top quark production processes constitute the pretraining dataset.

To test the pretrained model, we further generated four processes including three beyond Standard Model (SM) processes: a SM $t\bar{t}H$ production where the Higgs boson decays exclusively to a pair of photons, a $t\bar{t}H$ production with the Higgs boson decaying to a pair of photons, where the top-Yukawa coupling is CP-odd, implemented using the Higgs Characterization model~\cite{Artoisenet:2013puc}, the production of a pair of superpartners of the top quark (s-top) using the Minimal Supersymmetric Standard Model (MSSM)~\cite{PhysRevD.41.3464, ALLANACH20098}, and flavor changing neutral current (FCNC) processes~\cite{PhysRevD.91.034024, PhysRevD.91.074017}. For the s-top process, we simulate the production of heavier s-top pairs ($t_2\bar{t_2}$), where each heavier s-top (mass 582 GeV) decays into a lighter s-top ($t_1$ or $\bar{t_1}$, mass 400 GeV) and a Higgs boson. The FCNC process involves $t\bar{t}$ production where one top quark decays to a Higgs boson and a light quark. We generate 10 million events for each process, except for $tHq$ and $t\bar{t}t\bar{t}$, where 5 million events were produced.

In all simulation samples, the center of mass energy of the proton-proton collision is set to 13 TeV. The Higgs boson, top quarks, and vector bosons are set to decay inclusively (except the $t\bar{t}H \rightarrow \gamma\gamma$ samples), with MadSpin~\cite{Artoisenet:2012st} handling the decays of top quarks and W bosons. The generated events are processed through Pythia 8.235~\cite{Sjostrand:2014zea} for parton showering and heavy particle decays, followed by Delphes 3.4.2~\cite{deFavereau:2013fsa} configured to emulate the ATLAS detector~\cite{ATLAS_2008} for fast detector simulation.

The detector-level object selection criteria are defined to align with typical experimental conditions. Photons are required to have transverse momentum $p_T \geq 20$~GeV and pseudorapidity $|\eta| \leq 2.37$, excluding the electromagnetic calorimeter crack region ($1.37 < |\eta| < 1.52$). Electrons must have $p_T \geq 10$~GeV and $|\eta| \leq 2.47$ (excluding the same crack region), while muons are selected with $p_T \geq 10$~GeV and $|\eta| \leq 2.7$. Jets are reconstructed using the anti-$k_t$ algorithm~\cite{Cacciari:2008gp} with radius parameter $\Delta R=0.4$, where $\Delta R$ is defined as $\sqrt{\Delta\eta ^2 + \Delta\phi^2}$, with $\Delta\eta$ being the difference in pseudorapidity and $\Delta\phi$ the difference in azimuthal angle. Jets must satisfy $p_T \geq 25$~GeV and $|\eta| \leq 2.5$. To avoid double-counting, jets are removed if they are within $\Delta R < 0.4$ of a photon or lepton. The identification of jets originating from b-quark decays (b-tagging) is performed by matching jets within $\Delta R = 0.4$ of a b-quark, with efficiency corrections applied to match the performance of the ATLAS experiment's b-tagging algorithm~\cite{ATLAS:2019bwq}.

In addition to the simulated samples described above, we make use of the ATLAS Open Data release at 13 TeV \cite{ATLASOpenData2025}, which provides proton-proton collision data collected by the ATLAS detector during 2015 and 2016, corresponding to an integrated luminosity of 36 $\text{fb}^{-1}$. Unlike the Delphes-based fast simulation used in our pretraining and other fine-tuning samples, the ATLAS Open Data are processed through the full ATLAS detector simulation and reconstruction chain, making this dataset a stringent test of the generalizability of our pretrained models across different simulation frameworks. We used two event collections from this release: the \texttt{GamGam} collection, which requires at least two photons with $p_T \geq 25~\text{GeV}$ and is enriched in Higgs boson decays to diphotons, containing gluon fusion ($ggF$), vector boson fusion ($VBF$), and associated Higgs production ($WH$, $ZH$, and $t\bar{t}H$) processes; and the \texttt{1LMET30} collection, which requires at least one lepton with $p_T \geq 7~\text{GeV}$ and missing transverse energy of at least $30~\text{GeV}$, and is enriched with leptonically-decaying W bosons, containing triboson production processes ($WWW$, $ZZZ$, and other mixed triboson production $WWZ$ and $WZZ$).

The ATLAS Open Data triboson samples used in this analysis differ from our Delphes-based samples both at the event-generation and detector-simulation levels. Whereas our pretraining and Delphes fine-tuning samples are generated with MadGraph@NLO interfaced to Pythia and processed with Delphes fast detector simulation~\cite{Alwall:2014hca,Sjostrand:2014zea,deFavereau:2013fsa}, the ATLAS triboson samples used here are generated with Sherpa~2.2.2 and processed through the ATLAS detector simulation and reconstruction chain, with the detector response modeled using Geant4~\cite{Bothmann:2019yzt,ATLAS:2010wqa,Agostinelli:2002hh}. Therefore, the ATLAS Open Data triboson task probes not only a different detector simulation and reconstruction chain, but also a different event-generation and matrix-element setup from the MadGraph+Pythia+Delphes pipeline used for the Delphes samples.

\section{Methods}
\subsection{Overview}

We present a methodology for developing and evaluating a foundation model for particle collision event analysis. The approach centers on pretraining a Graph Neural Network (GNN) architecture using a comprehensive dataset that spans multiple physics tasks, enabling the model to learn robust and transferable features. For task-specific applications, we employ a fine-tuning strategy that combines output layer adaptation with carefully calibrated learning rates for updating the pretrained parameters.

Given the prevalence of classification problems in particle physics data analysis, we evaluate the model's efficacy through a systematic assessment across five binary classification tasks and two multiclass classification tasks:
\begin{itemize}
   \item $t\bar{t}H(\rightarrow \gamma\gamma)$ with CP-even versus CP-odd t-H interaction (Delphes)
   \item $t\bar{t}$ with FCNC top quark decays versus $tHq$ processes (Delphes)
   \item $t\bar{t}W$ versus $t\bar{t}t$ processes (Delphes)
   \item s-top pair production with Higgs bosons in the decay chain versus $t\bar{t}H$ processes (Delphes)
   \item $WH$ versus $ZH$ production modes (Delphes)
   \item 5-class multiclass classification of Higgs production modes where the Higgs decays exclusively to two photons: $t\bar{t}H$, $ggF$, $VBF$, $WH$, and $ZH$ (ATLAS Open Data)
   \item 3-class multiclass classification of triboson events: $WWW$, $ZZZ$ and Mixed ($WWZ$ and $WZZ$) (ATLAS Open Data)
\end{itemize}

Our evaluation metrics encompass classification performance, computational efficiency, and model interpretability. The investigation extends to analyzing the model's scaling behavior with respect to training dataset size, benchmarked against models trained without pretraining. Although we explored transfer learning through parameter freezing of pretrained layers, this approach did not yield performance improvements, leading us to focus our detailed analysis on fine-tuning strategies.

This methodological framework demonstrates the potential of foundation models to enhance the efficiency of particle physics analyses while improving task-specific performance, offering a promising direction for future high-energy physics research.

\subsection{GNN Architecture}
We implement a Graph Neural Network (GNN) architecture that naturally accommodates the point-cloud structure of particle physics data, employing the \texttt{DGL} framework with a \texttt{PyTorch} backend \cite{DGL, PyTorch}. The GNN naturally handles graphs of varying node and edge counts through the message-passing framework, without requiring padding or truncation, making it well suited to collision events where the number of reconstructed objects varies from event to event. A fully connected graph is constructed for each event, with nodes corresponding to reconstructed jets, electrons, muons, photons, and $\vec{E}_T^{\text{miss}}$. The features of each node include the four-momentum $(p_T,\eta,\phi,E)$ of the object with a massless assumption ($E=p_T\cosh\eta$), the b-tagging label (for jets), the charge (for leptons), and an integer labeling the type of object represented by the node. We use a placeholder value of 0 for features which are not defined for every node type such as the b-jet tag, lepton charge, or the pseudorapidity of $\vec{E}_T^{\text{miss}}$. An explicit masking mechanism was not tested in this work and represents a potential avenue for future improvement. We assign the angular distances ($\Delta \eta, \Delta \phi, \Delta R$) as edge features and the number of nodes $N$ in the graph as a global feature. We denote the node features $\{\vec x_i\}$, edge features $\{\vec y_{ij}\}$, and global features $\{\vec z\}$.

The GNN model is based on the graph network architecture described in \cite{graph_nets} using simple multilayer perceptron (MLP) feature functions and summation aggregation. The model is comprised of three primary components: an encoder, the graph network, and a decoder. In the encoder, three MLPs embed the nodes, edges, and global features into a latent space of dimension 64. The graph network block, which is designed to facilitate message passing between different domains of the graph, performs an edge update $f_e$, followed by a node update $f_n$, and finally a global update $f_g$, all defined below. The inputs to each update MLP are concatenated.

\[\vec {y'}_{ij} = f_e\left(\{\vec x_k\},\vec y_{ij},\vec z\right) = \mathrm{MLP}\left(\vec x_i,\vec x_j,\vec y_{ij},\vec z\right)\]

\[\vec{x'}_{i} = f_n\left(\vec x_i,\{\vec{y'}_{jk}\},\vec z\right) = \mathrm{MLP}\left(\vec x_i,\sum_j\vec{y'}_{ij},\vec z\right)\]

\[\vec{z'} = f_g\left(\{\vec{x'}_i\},\{\vec{y'}_{ij}\},\vec z\right) = \mathrm{MLP}\left(\sum_i\vec{x'}_i,\sum_{i,j}\vec{y'}_{ij},\vec z\right)\]

This graph block is iterated four times with the same update MLPs. Finally, the global features are passed through a decoder MLP and a final linear layer to produce the desired model outputs. Each MLP consists of 4 linear layers, each with an output width of 64, with the \texttt{ReLU} activation function. The output of the MLP is then passed through a \texttt{LayerNorm} layer\cite{LayerNorm}. The total number of trainable parameters in this model is about 400,000.

As a performance benchmark, a baseline GNN model is trained from scratch for each classification task. The Adam optimizer is used, with an initial learning rate of $10^{-4}$ and an exponential decay schedule given by $\mathrm{LR}(x) = \mathrm{LR}_{\mathrm{initial}}\cdot(0.99)^x$, where $x$ denotes the epoch number.

The baseline hyperparameters of the GNN are summarized in Table~\ref{tab:hyperparameters}.

\begin{table}[h]
\centering
\caption{Baseline GNN Hyperparameters.}
\label{tab:hyperparameters}
\begin{tabular}{lc}
\hline
\textbf{Hyperparameter} & \textbf{Value} \\
\hline
Hidden Layer Size & 64 \\
MLP Layers & 4 \\
Graph Processing Steps & 4 \\
Batch Size & 1024 \\
Optimizer & Adam \\
Initial Learning Rate & $10^{-4}$ \\
Learning Rate Decay Factor & $0.99$ \\

\hline
\end{tabular}
\end{table}

A total of 400 GPU hours on a single NVIDIA A100 were invested for 
systematic hyperparameter optimization, with full results reported in 
Appendix~\ref{app:hyperparameter}. We perform a sweep over batch size, 
learning rate, MLP depth and width, and hidden dimension. These results 
show that the GNN architecture exhibits stability across hyperparameter 
configurations: excluding a single outlier, variations in the test AUC 
remain within $\pm 0.2\%$.

\subsection{Pretraining Strategy}

We explore two complementary pretraining approaches to develop robust representations of collision events: (1) multiclass classification, which trains the model to distinguish between different physics processes, and (2) multilabel classification, which predicts the existence and kinematics of heavy particles with prompt decays. The pretraining dataset consists of approximately 120 million events, evenly distributed across 12 distinct physics processes, including all major Higgs boson production mechanisms and top quark processes as described in Section~\ref{sec:data}. This large-scale pretraining effort was conducted on the Perlmutter supercomputer at NERSC.

\subsubsection{Multiclass Classification}
For Monte Carlo simulated events, the underlying physics process that generated each event is known precisely, providing natural labels for supervised learning. However, the challenge lies in the complexity of collision events: different physics processes can produce similar kinematics and event topologies, particularly in certain regions of phase space. No single observable can unambiguously identify the underlying process. By training the model to distinguish between 12 different processes simultaneously, we challenge it to learn subtle differences in kinematics and topology that collectively characterize each process. The model is trained using categorical cross entropy as the loss function. The output layer of the multiclass classification model has 832 trainable parameters.

\subsubsection{Multilabel Classification}
This approach combines both classification and regression tasks to characterize collision events. For discrete properties like particle presence in specific kinematic regions, we employ classification labels with binary cross-entropy loss. For continuous quantities like particle multiplicities, we use regression labels with mean-squared error loss. This hybrid approach enables the model to learn both categorical and continuous aspects of the physics processes simultaneously.

We develop a comprehensive set of 41 labels that capture both particle multiplicities and kinematic properties. This approach increases prediction granularity and enhances model interpretability. By training the model to predict event kinematics rather than event identification, we create a task-independent framework that can potentially generalize better to novel scenarios not seen during pretraining.

The particle multiplicity labels count the number of Higgs bosons ($n_{\text{higgs}}$), top quarks ($n_{\text{tops}}$), vector bosons ($n_V$), $W$ bosons ($n_W$), and $Z$ bosons ($n_Z$). The kinematic labels characterize the transverse momentum ($p_T$), pseudorapidity ($\eta$), and azimuthal angle ($\phi$) of Higgs bosons and top quarks through binned classifications.

For Higgs bosons, $p_T$ is categorized into three ranges: (0, 30) GeV, (30, 200) GeV, and (200, $\infty$) GeV, with the upper range particularly sensitive to potential BSM effects. Similarly, both leading and subleading top quarks have $p_T$ classifications spanning (0, 30) GeV, (30, 300) GeV, and (300, $\infty$) GeV. When no particle exists within a specific $p_T$ range, the corresponding label is set to $[0, 0, 0]$. For all particles, $\eta$ measurements are divided into 4 bins with boundaries at $[-1.5, 0, 1.5]$, while $\phi$ measurements use 4 bins with boundaries at $[-\frac{\pi}{2}, 0, \frac{\pi}{2}]$. As with $p_T$, both $\eta$ and $\phi$ labels default to $[0, 0, 0, 0]$ in the absence of a particle. This comprehensive labeling schema enables fine-grained learning of kinematic distributions and particle multiplicities, essential for characterizing complex collision events.

The loss function combines individual losses from all 41 labels through weighted averaging. Binary cross-entropy is applied to classification labels, while mean-squared error is used for regression labels. The model generates predictions for all labels simultaneously, with individual losses calculated according to their respective types. The final loss is computed as an equally-weighted average across all labels, with weights set to 1 to ensure uniform contribution to the optimization process. The output layer of the multilabel model has 2,688 trainable parameters.

\subsubsection{Pretraining}
During pretraining, the initial learning rate is $10^{-4}$, and the learning rate decays by 1\% each epoch following the exponential decay function $LR(x) = 10^{-4}\cdot(0.99)^x$, where $x$ is the number of epochs. The model is trained until the training loss shows no improvement for $5$ consecutive epochs, indicating convergence in the training. The epoch with the maximal validation AUC is then selected for fine-tuning.

\subsection{Fine-tuning Methodology}

For downstream tasks, we adjust the model architecture for fine-tuning by replacing the original output layer (final linear layer) with a newly initialized linear layer while retaining the pretrained weights for all other layers. This modification allows the model to specialize in the specific downstream task while leveraging the general features learned during pretraining.

The fine-tuning process begins with distinct learning rate setups for different parts of the model. The newly initialized linear layer is trained with an initial learning rate of $10^{-4}$, matching the rate used for models trained from scratch. Meanwhile, the pretrained layers are fine-tuned more cautiously with a lower initial learning rate of $10^{-5}$. This approach ensures that the pretrained layers adapt gradually without losing their general features, while the new layer learns effectively from scratch. Both learning rates decay over time following the same exponential decay function, $LR(x) = LR_{initial} \cdot (0.99)^x$, to promote stable convergence as training progresses.

We also evaluated a transfer learning setup in which either the decoder MLP or the final linear layer was replaced with a newly initialized component. During this process, all other model parameters remained frozen, leveraging the pretrained features without further updating them. However, we did not observe performance improvements using the transfer learning setup. Consequently, we focus on reporting results obtained with the fine-tuning approach.

\subsection{Performance Evaluation}

We assess model performance using two figures of merit: the classification accuracy and the Area Under the Curve (AUC) of the Receiver Operating Characteristic (ROC) curve. For binary tasks, accuracy is computed after applying a threshold of 0.5 to the neural-network output score. For multiclass tasks, the predicted class is taken to be the class with the largest output score. Both metrics demonstrate consistent trends in our analysis.

We employ an ensemble training approach where 5 independent models are trained for each configuration with random weight initialization and random subsets of the training dataset. This enables us to evaluate both the model's sensitivity to initial parameters and to quantify uncertainties in their performance. All models are trained to the same stopping condition: $5$ consecutive epochs where there are no improvements to the training loss, and the epoch with maximal validation AUC is then selected for further performance evaluation using the held-out test set. To investigate how model performance scales with training data, we conduct training runs using sample sizes ranging from $10^3$ to $10^7$ events per class ($10^3$, $10^4$, $10^5$, $10^6$, and $10^7$) for each model setup: the from-scratch baseline and models fine-tuned from multiclass or multilabel pretrained models. For the $10^7$ Delphes case and the ATLAS Open Data tasks, only the initialization was randomized due to dataset size limitations.

Additional performance metrics, including F1, precision, and recall, are reported in Appendix~\ref{app:performance} and show trends consistent with the accuracy and AUC.

\begin{table*}[t]
\centering
\begin{tabular}{c c c ccccc}
    \hline
    Name of Task & Metric & Pretraining & \multicolumn{5}{c}{Sample Size} \\ 
    \cline{4-8}
     & & & $10^3$ & $10^4$ & $10^5$ & $10^6$ & $10^7$ \\
    \hline\hline
    \multirow{6}{*}{$t\bar{t}H(\rightarrow\gamma\gamma)$ CP-even vs CP-odd}
    & \multirow{3}{*}{Accuracy (\%)} 
    & Baseline      & 56.4 $\pm$ 1.0 & 62.3 $\pm$ 0.1 & 64.2 $\pm$ 0.1 & 65.7 $\pm$ 0.0 & 66.2 $\pm$ 0.0 \\
    & & Multiclass   & +2.9 $\pm$ 1.1 & +2.0 $\pm$ 0.1 & +1.0 $\pm$ 0.1 & +0.1 $\pm$ 0.0 & -0.0 $\pm$ 0.0 \\
    & & Multilabel   & +1.3 $\pm$ 1.2 & +1.0 $\pm$ 0.2 & +0.6 $\pm$ 0.1 & +0.0 $\pm$ 0.0 & -0.1 $\pm$ 0.0 \\
    \cline{2-8}
    & \multirow{3}{*}{AUC (x 100)} 
    & Baseline      & 59.1 $\pm$ 1.4 & 66.9 $\pm$ 0.2 & 69.5 $\pm$ 0.0 & 71.2 $\pm$ 0.0 & 71.9 $\pm$ 0.0 \\
    & & Multiclass   & +3.9 $\pm$ 1.4 & +2.5 $\pm$ 0.2 & +1.0 $\pm$ 0.0 & +0.1 $\pm$ 0.0 & -0.0 $\pm$ 0.0 \\
    & & Multilabel   & +0.6 $\pm$ 1.6 & +1.1 $\pm$ 0.2 & +0.6 $\pm$ 0.1 & -0.0 $\pm$ 0.0 & -0.1 $\pm$ 0.0 \\
    \hline
    \multirow{6}{*}{FCNC vs $tHq$}
    & \multirow{3}{*}{Accuracy (\%)} 
    & Baseline      & 62.2 $\pm$ 0.6 & 67.9 $\pm$ 0.4 & 68.0 $\pm$ 0.4 & 69.2 $\pm$ 0.2 & 67.7 $\pm$ 0.1 \\
    & & Multiclass   & +4.6 $\pm$ 0.8 & +0.7 $\pm$ 0.4 & +1.4 $\pm$ 0.5 & +0.4 $\pm$ 0.2 & -0.2 $\pm$ 0.1 \\
    & & Multilabel   & -1.9 $\pm$ 0.8 & -1.1 $\pm$ 0.5 & +1.0 $\pm$ 0.5 & -0.0 $\pm$ 0.3 & +0.0 $\pm$ 0.1 \\
    \cline{2-8}
    & \multirow{3}{*}{AUC (x 100)} 
    & Baseline      & 67.6 $\pm$ 0.5 & 72.1 $\pm$ 0.0 & 74.3 $\pm$ 0.0 & 75.6 $\pm$ 0.0 & 73.8 $\pm$ 0.0 \\
    & & Multiclass   & +2.5 $\pm$ 1.0 & +2.6 $\pm$ 0.1 & +1.4 $\pm$ 0.1 & +0.6 $\pm$ 0.0 & -0.0 $\pm$ 0.0 \\
    & & Multilabel   & -3.4 $\pm$ 1.0 & -0.3 $\pm$ 0.1 & +0.4 $\pm$ 0.0 & +0.0 $\pm$ 0.0 & -0.1 $\pm$ 0.0 \\
    \hline
    \multirow{6}{*}{$t\bar{t}W$ vs $t\bar{t}t$}
    & \multirow{3}{*}{Accuracy (\%)} 
    & Baseline      & 75.7 $\pm$ 0.1 & 77.5 $\pm$ 0.1 & 79.0 $\pm$ 0.0 & 79.8 $\pm$ 0.0 & 80.3 $\pm$ 0.0 \\
    & & Multiclass   & +2.7 $\pm$ 0.2 & +2.2 $\pm$ 0.1 & +1.0 $\pm$ 0.0 & +0.4 $\pm$ 0.0 & +0.0 $\pm$ 0.0 \\
    & & Multilabel   & +1.6 $\pm$ 0.1 & +0.5 $\pm$ 0.2 & +0.3 $\pm$ 0.0 & +0.0 $\pm$ 0.0 & -0.1 $\pm$ 0.0 \\
    \cline{2-8}
    & \multirow{3}{*}{AUC (x 100)} 
    & Baseline      & 83.3 $\pm$ 0.1 & 85.5 $\pm$ 0.1 & 87.2 $\pm$ 0.0 & 88.0 $\pm$ 0.0 & 88.5 $\pm$ 0.0 \\
    & & Multiclass   & +2.9 $\pm$ 0.2 & +2.2 $\pm$ 0.1 & +1.0 $\pm$ 0.0 & +0.3 $\pm$ 0.0 & +0.0 $\pm$ 0.0 \\
    & & Multilabel   & +1.1 $\pm$ 0.2 & +0.9 $\pm$ 0.1 & +0.4 $\pm$ 0.0 & +0.0 $\pm$ 0.0 & -0.1 $\pm$ 0.0 \\
    \hline
    \multirow{6}{*}{s-top vs $t\bar{t}H$}
    & \multirow{3}{*}{Accuracy (\%)} 
    & Baseline      & 82.9 $\pm$ 0.2 & 86.3 $\pm$ 0.1 & 87.5 $\pm$ 0.1 & 88.5 $\pm$ 0.0 & 88.8 $\pm$ 0.0 \\
    & & Multiclass   & +0.4 $\pm$ 0.3 & +1.6 $\pm$ 0.1 & +0.9 $\pm$ 0.1 & +0.3 $\pm$ 0.0 & +0.0 $\pm$ 0.0 \\
    & & Multilabel   & +1.9 $\pm$ 0.2 & +0.8 $\pm$ 0.1 & +0.5 $\pm$ 0.1 & +0.0 $\pm$ 0.0 & -0.0 $\pm$ 0.0 \\
    \cline{2-8}
    & \multirow{3}{*}{AUC (x 100)} 
    & Baseline      & 90.3 $\pm$ 0.2 & 93.7 $\pm$ 0.0 & 94.7 $\pm$ 0.0 & 95.4 $\pm$ 0.0 & 95.6 $\pm$ 0.0 \\
    & & Multiclass   & +0.4 $\pm$ 0.2 & +1.2 $\pm$ 0.0 & +0.6 $\pm$ 0.0 & +0.2 $\pm$ 0.0 & +0.0 $\pm$ 0.0 \\
    & & Multilabel   & +0.9 $\pm$ 0.2 & +0.6 $\pm$ 0.1 & +0.3 $\pm$ 0.0 & +0.0 $\pm$ 0.0 & -0.0 $\pm$ 0.0 \\
    \hline
    \multirow{6}{*}{$WH$ vs $ZH$}
    & \multirow{3}{*}{Accuracy (\%)} 
    & Baseline      & 51.2 $\pm$ 0.1 & 54.0 $\pm$ 0.1 & 55.8 $\pm$ 0.1 & 57.6 $\pm$ 0.0 & 58.0 $\pm$ 0.0 \\
    & & Multiclass   & +2.8 $\pm$ 0.2 & +2.8 $\pm$ 0.1 & +1.7 $\pm$ 0.1 & +0.4 $\pm$ 0.0 & +0.1 $\pm$ 0.0 \\
    & & Multilabel   & -0.4 $\pm$ 0.2 & -0.5 $\pm$ 0.2 & +0.3 $\pm$ 0.1 & +0.0 $\pm$ 0.0 & -0.1 $\pm$ 0.0 \\
    \cline{2-8}
    & \multirow{3}{*}{AUC (x 100)} 
    & Baseline      & 51.8 $\pm$ 0.1 & 56.1 $\pm$ 0.2 & 59.5 $\pm$ 0.0 & 62.1 $\pm$ 0.0 & 62.7 $\pm$ 0.0 \\
    & & Multiclass   & +4.6 $\pm$ 0.2 & +4.8 $\pm$ 0.2 & +2.5 $\pm$ 0.0 & +0.5 $\pm$ 0.0 & +0.1 $\pm$ 0.0 \\
    & & Multilabel   & -0.9 $\pm$ 0.2 & -0.3 $\pm$ 0.3 & +0.4 $\pm$ 0.2 & +0.0 $\pm$ 0.0 & -0.1 $\pm$ 0.0 \\
    \hline
\end{tabular}
\caption{Accuracy and AUC of the baseline model versus the improvement due to fine-tuning from various pretraining tasks, computed using a held-out test dataset consisting of one million events per class. Model checkpoints are selected using a separate validation dataset also consisting of one million events per class. AUC values are multiplied by 100 for display purposes. These metrics are averaged over 5 independently trained models with randomly initialized weights and trained on a random subset of the data. One exception is the $10^7$ training where all models use the same dataset due to limitations on our dataset size. The random subsets are allowed to overlap, but this overlap should be very minimal because all models take an independent random subset of $10^7$ events. The errors are the propagated errors (root sum of squares) of the standard deviation of accuracies and AUCs for each model.}
\label{tab:results}
\end{table*}

\section{Results}
\subsection{Classification Performance}

In general, the fine-tuned pretrained model achieves at least the same level of classification performance as the baseline model. Notably, there are significant improvements, particularly when the sample size is small, ranging from $10^3$ to $10^5$ events. In some cases, the accuracy (AUC) improvements exceed 4 percentage points (2.5 AUC points), demonstrating that pretrained models provide a strong initial representation that compensates for limited data. The numerical values of the improvements in accuracy may not fully capture the impact on the sensitivity of the measurements for which the neural network classifier is used, and the improvement in physics analysis sensitivity is likely to be greater.

As the training sample size grows to $10^6$ and eventually $10^7$ events, the added benefit of pretraining diminishes. With abundant data, models trained from scratch approach or even match the accuracy of fine-tuned pretrained models. This suggests that large datasets enable effective learning from scratch, rendering the advantage of pretraining negligible in such scenarios.

Although both pretraining approaches offer benefits, multiclass pretraining tends to provide more consistent improvements across tasks, especially in the low-data regime. In contrast, multilabel pretraining can sometimes lead to neutral or even slightly negative effects for certain tasks and data sizes. This highlights the importance of the pretraining task design, as the similarity between pretraining and fine-tuning tasks in the multiclass approach appears to yield better-aligned representations.

We note that training sample sizes of $10^3$ to $10^5$ events per class reflect realistic operating conditions in HEP analyses. Searches for new physics involve simulated signal samples that are expensive to generate due to stringent physics selections, and standard analysis workflows further reduce per-category training statistics by partitioning events into multiple exclusive phase space regions, each requiring a dedicated classifier. The observed performance gains in this regime are therefore of direct practical relevance.

Finally, the spread of accuracy across the five tasks for the baseline model is quite large, offering a robust test of fine-tuning across tasks of varying difficulty. The consistent observation of these trends across tasks confirms the reliability and robustness of the findings.

\subsection{Generalizability to ATLAS Open Data}

\begin{table*}[t]
    \centering
    \begin{tabular}{c c c c c}
        \hline
        Task & Dataset Size & Pretraining & Accuracy (\%) & AUC ($\times$100) \\
        \hline\hline
        \multirow{3}{*}{Higgs Production}
        & \multirow{3}{*}{2,500,000}
        & Baseline    & $71.8 \pm 0.2$ & $91.2 \pm 0.0$ \\
        & & Multiclass  & $-0.1 \pm 0.2$ & $+0.1 \pm 0.0$ \\
        & & Multilabel  & $-1.6 \pm 0.2$ & $-0.6 \pm 0.0$ \\
        \hline
        \multirow{3}{*}{Triboson}
        & \multirow{3}{*}{300,000}
        & Baseline    & $54.8 \pm 2.3$ & $72.9 \pm 3.3$ \\
        & & Multiclass  & $+3.7 \pm 2.3$ & $+4.6 \pm 3.3$ \\
        & & Multilabel  & $-1.7 \pm 2.5$ & $-1.2 \pm 3.8$ \\
        \hline
    \end{tabular}
    \caption{Classification accuracy and AUC for the ATLAS Open Data tasks, computed using a held-out test dataset (10\%). Model checkpoints are selected using a separate validation dataset (10\%), with the rest being used for training (80\%). The Multiclass and Multilabel rows report the absolute improvement over the baseline. AUC values are multiplied by 100 for display. Uncertainties are the standard deviation across 5 independently trained models. No training sample size scaling study was performed.}
    \label{tab:opendata_results}
\end{table*}

We evaluate the generalizability of the pretrained models on two multiclass classification tasks using the ATLAS Open Data, described in Section \ref{sec:data}. These tasks are designed to probe whether representations learned on Delphes-simulated events transfer to data processed through the full ATLAS detector simulation and reconstruction chain, which differs significantly in its treatment of detector effects, pile-up, and object reconstruction. 

The first task uses the \texttt{GamGam} collection to classify five Higgs boson production modes ($ggF$, $VBF$, $WH$, $ZH$, and $t\bar{t}H$) where the Higgs boson decays exclusively to a pair of photons. Distinguishing between these production modes is of direct physics interest, as each is sensitive to different couplings of the Higgs boson to other SM particles. The $VBF$ and $t\bar{t}H$ production modes in particular are important probes of the Higgs boson's couplings to vector bosons and top quarks respectively, and improving their separation from other production modes has direct implications for coupling measurements at the LHC.

The second task uses the \texttt{1LMET30} collection to classify three triboson production processes: $WWW$, $ZZZ$, and Mixed ($WWZ$, and $WZZ$). Triboson production is a rare SM process that provides a direct probe of quartic gauge boson couplings, and any deviation from SM predictions would be a strong indication of new physics. The relatively small dataset of $300,000$ events reflect the rarity of these processes, making this task a particularly stringent test of the pretrained model's ability to compensate for limited training data.

The results are summarized in Table \ref{tab:opendata_results}. For the Higgs production task, multiclass pretraining yields no improvement with $-0.1 \pm 0.2$ percentage points in accuracy and a small increase of $+0.1 \pm 0.0$ AUC points over the baseline, while multilabel pretraining leads to a slight degradation in both metrics. For the triboson task, multiclass pretraining provides a more substantial improvement of $+3.7 \pm 2.3$ percentage points in accuracy and $+4.6 \pm 3.3$ AUC points, consistent with the trend observed in the Delphes-based evaluation where pretraining benefits are largest when the training dataset is small. Once again, the multilabel pretraining again performs poorly with a slight degradation in both metrics.

The consistent improvement from multiclass pretraining across both simulation frameworks suggests that the pretrained model learns physics-driven representations that are robust to differences in detector simulation. The failure of multilabel pretraining in both the Delphes and ATLAS Open Data evaluations points to a systematic limitation of this pretraining strategy, possibly reflecting a misalignment between the multilabel pretraining objective and the fine-tuning tasks.

The poor performance of the multilabel pretraining in the low statistics regime for both the Delphes and ATLAS Open Data evaluations point to a systematic limitation of this pretraining strategy. We interpret this behavior as evidence that not all physically motivated supervised pretraining objectives produce equally transferable representations. In the multilabel setup, the model is explicitly optimized to predict a fixed set of hand-designed labels, including heavy-particle multiplicities and coarse kinematic properties. Although these labels are physically meaningful, they impose a specific structure on the learned representation: the network is encouraged to organize events according to predefined particle-level quantities rather than to discover the event-level features most useful for downstream classification. This can lead to negative transfer when the fine-tuning task depends on information that is weakly captured by, or not well aligned with, the multilabel targets. For example, tasks such as $WH$ versus $ZH$, FCNC versus $tHq$, and the ATLAS Open Data triboson classification may depend on subtle correlations among reconstructed objects, detector-level features, or global event topology that are not fully captured by the chosen multilabel objective. In contrast, the multiclass pretraining objective directly requires the model to discriminate among complete physics processes, allowing it to learn whichever combination of object kinematics, multiplicities, and graph-level correlations is most useful for event separation. These results suggest that future pretraining objectives should avoid overly prescriptive auxiliary labels unless those labels are closely aligned with the intended downstream tasks. More flexible event-level objectives, or objectives based on discrimination among complete event topologies, may produce representations with better transferability.

\subsection{Model Interpretability}

We aim to understand whether pretrained and baseline models learn the same underlying representations. To investigate this, we perform a layer-wise CKA analysis across three model comparisons: (1) the multiclass pretrained model \textit{before} fine-tuning versus the best-performing baseline, (2) the best fine-tuned model versus the best-performing baseline, and (3) the best fine-tuned model versus the pretrained model before fine-tuning. Together, these three comparisons allow us to disentangle the effects of pretraining initialization from those of fine-tuning adaptation and to understand how representations evolve during fine-tuning.

We use Centered Kernel Alignment (CKA) ~\cite{DBLP:journals/corr/abs-1905-00414} to quantify the similarity between the internal representations of neural networks by comparing their feature matrices in a manner that is invariant to scaling, rotation, and alignment. This invariance makes CKA particularly effective for studying relationships between network layers, even across networks trained from varying initialization. 

To provide an intuitive understanding of CKA values, we construct a table of the CKA scores for various transformations performed on a set of dummy data, shown in Table \ref{tab:cka_dummy_data}. 

\begin{itemize}
    \item $A$: randomly initialized matrix with shape (1000, 64), following a normal distribution ($\sigma = 1, \mu=0$)
     \item $B$: matrix with shape (1000, 64) constructed via various transformations performed on $A$
     \item $Noise$: randomly initialized noise matrix with shape (1000, 64), following a normal distribution ($\sigma = 1, \mu=0$)
\end{itemize}

\begin{table}[h]
    \centering
    \begin{tabular}{l|c}
        \hline
        Dataset & CKA Score\\
        \hline \hline
        $A, B = A$ & 1.00 \\
        $A, B =$ permutation on columns of $A$ & 1.00 \\
        $A, B = A + \mathrm{Noise}(0.1)$ & 0.99 \\ 
        $A, B = A + \mathrm{Noise}(0.5)$ & 0.80 \\ 
        $A, B = A + \mathrm{Noise}(0.75)$ & 0.77 \\ 
        $A, B = A\cdot \mathrm{Noise}(1)$ (Linear Transformation) & 0.76 \\
        $A, B = A + \mathrm{Noise}(1)$ & 0.69 \\ 
        $A, B = A + \mathrm{Noise}(2)$ & 0.51 \\ 
        $A, B = A + \mathrm{Noise}(5)$ & 0.39 \\ 
        \hline
    \end{tabular}
    \caption{CKA scores for a dummy dataset $A$ and $B$, where $B$ is created via various transformations performed on $A$. This provides an interpretable reference scale: values above $\sim 0.8$ indicate highly similar representations, values around $\sim 0.5$ are comparable to adding noise of approximately 2 standard deviations, and values below $\sim 0.4$ indicate substantially different representations.}
    \label{tab:cka_dummy_data}
\end{table}

We train ensembles of models for each training task to observe how the CKA score changes due to the random initialization. The CKA score between two models is then defined as:
\begin{align}
    CKA(A, B) = \frac{1}{n^2}\sum_i^n \sum_j^nCKA(A_i, B_j)
\end{align}
where $A_i$ is the representation learned by the $i^{th}$ model in an ensemble with $n$ total models. The uncertainty in CKA is the standard deviation of $CKA(A_i, B_j)$.

\begin{figure*}[t]
    \centering
    \includegraphics[width=\textwidth]{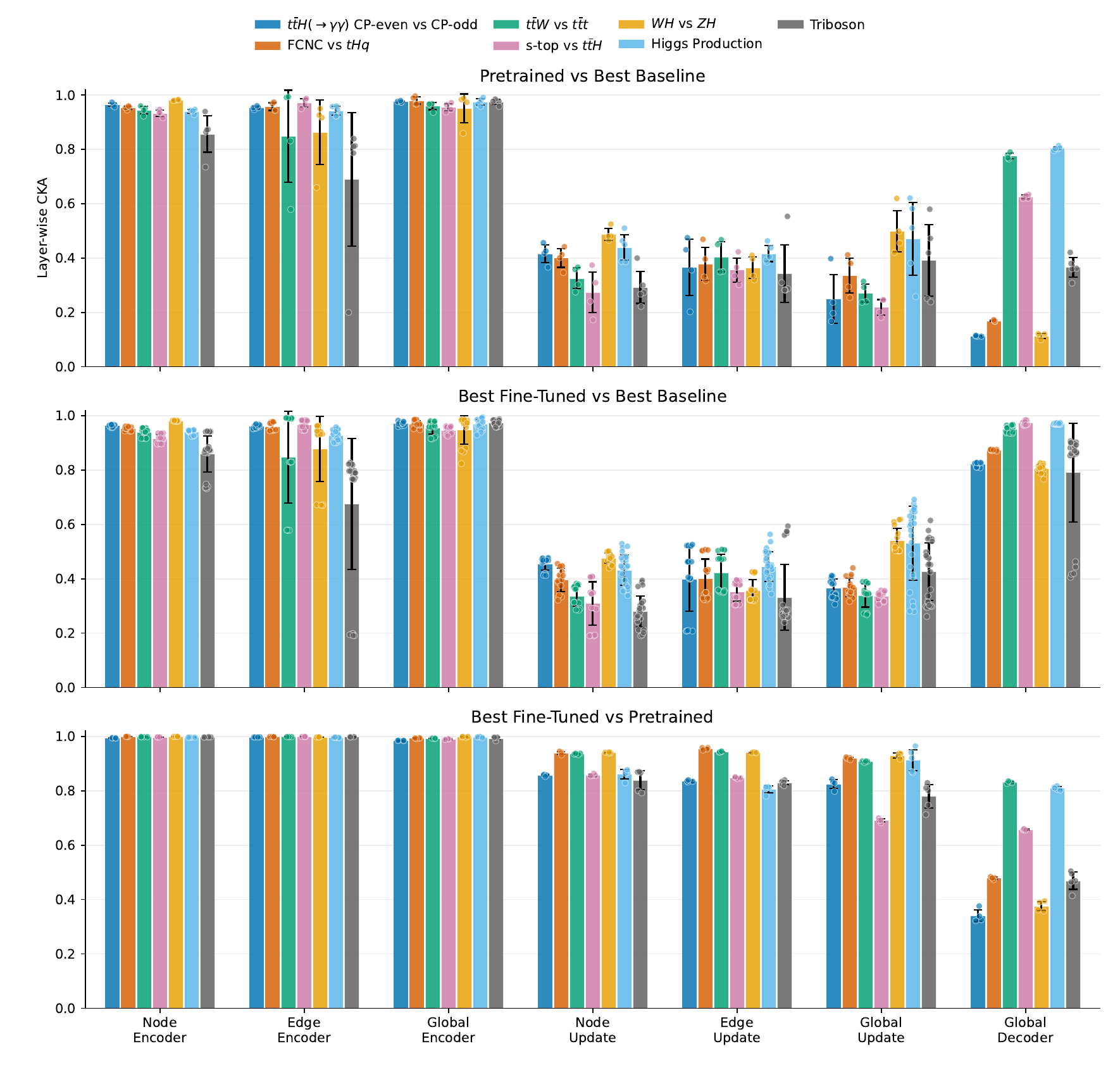}
    \caption{Layer-wise CKA similarity across three model comparisons: (top) multiclass pretrained model before fine-tuning versus the best-performing baseline, (middle) best fine-tuned model versus the best-performing baseline, and (bottom) best fine-tuned model versus the pretrained model before fine-tuning. Each panel shows the CKA similarity at each stage of the GNN architecture across all downstream tasks. For each model pair, the CKA is computed using a fixed evaluation sample of 1024 events selected from the held-out test set, and the reported CKA value is averaged over all pairwise combinations of the five independently trained models in each ensemble. Error bars represent the standard deviation across these pairwise model comparisons. The evaluation sample size was chosen for computational tractability, since the layer-wise model comparison is expensive to evaluate; although this sample is small compared to the full dataset, the small uncertainties indicate that the observed representation-level trends are stable across independently trained models.}
    \label{fig:cka_layerwise}
\end{figure*}

The GNN architecture consists of three initial encoders (Node Encoder, Edge Encoder, Global Encoder), followed by graph processing stages applied iteratively in a message passing loop (Node Update, Edge Update, Global Update), and a final Global Decoder. The layer-wise CKA similarity for all three comparisons is shown in Figure~\ref{fig:cka_layerwise} and reveals three distinct and complementary perspectives on how pretraining and fine-tuning shape model representations.

\subsubsection*{Pretrained vs Best Baseline}

The first comparison examines the representational similarity between the pretrained model, before any fine-tuning, and the best-performing baseline trained directly on each downstream task. Three distinct regimes are visible. In the encoding stages (Node Encoder, Edge Encoder, Global Encoder), CKA values are consistently high across all tasks, in the range of $0.9$--$1.0$, indicating that the pretrained model has already developed low-level feature representations that closely resemble those of the best baseline, despite never having been trained on the downstream task, suggesting that pretraining on a diverse set of physics processes produces encoder representations that are nearly identical to those learned by a model fully optimized for the target classification. 

In the message passing stages (Node Update, Edge Update, Global Update), the similarity drops dramatically to values in the range of $0.2$--$0.5$ across all tasks. Referencing Table~\ref{tab:cka_dummy_data}, these values are comparable to adding noise of approximately 2 standard deviations, indicating that the pretrained and baseline models process and aggregate graph information through fundamentally different computational pathways. This suggests that the diverse pretraining objectives lead the model to develop more general intermediate representations that differ substantially from those optimized purely for the downstream task.

At the Global Decoder, the similarity is task-dependent and correlates with task 
difficulty. For the harder tasks ($t\bar{t}H(\rightarrow\gamma\gamma)$ CP-even vs CP-odd, FCNC vs $tHq$, and $WH$ vs $ZH$) the decoder representations remain 
substantially different from the baseline, with CKA values as low as $0.1$--$0.2$. For the easier tasks ($t\bar{t}W$ vs $t\bar{t}t$ and s-top vs $t\bar{t}H$) the similarity partially recovers to $0.7$--$0.8$. This pattern suggests that for difficult tasks, the pretrained model's divergent intermediate representations encode inductive biases that are not accessible to the baseline model, which may explain the larger performance improvements observed for these tasks in the low-data regime.

\subsubsection*{Best Fine-Tuned vs Best Baseline}

The second comparison examines whether fine-tuning drives the pretrained model toward the same representational solution as the baseline, or whether it retains a distinct learned structure. The encoding stages again show high similarity in the range of $0.9$--$1.0$, consistent with the first comparison and confirming that these low-level representations are stable across all training strategies. The message passing stages remain similarly low, with CKA values of $0.2$--$0.5$, indicating that fine-tuning does not recover the baseline's intermediate representations. The fine-tuned model retains the structurally different message passing pathway of the pretrained model even after task-specific training.

The Global Decoder shows a marked improvement in similarity compared to the pretrained vs baseline comparison, with values consistently in the range of $0.8$--$1.0$ across nearly all tasks. This indicates that fine-tuning primarily acts on the final decoder stage, driving the model's output representations into close alignment with those of the best baseline regardless of task difficulty. The fine-tuned model therefore arrives at a similar decision boundary to the baseline through a fundamentally different intermediate computational pathway, which may explain why fine-tuned models can match or exceed baseline performance while maintaining distinct internal representations. 

\subsubsection*{Best Fine-Tuned vs Pretrained}

The third comparison directly measures how much fine-tuning modifies the pretrained representations. In the encoding stages, the CKA similarity is nearly $1.0$ across all tasks, confirming that fine-tuning leaves the low-level encoder representations essentially unchanged. This is consistent with the standard intuition that early layers in a pretrained network encode general features that do not need to be modified for downstream tasks.

In the message passing stages, the similarity remains high, in the range of $0.8$--$0.95$ for most tasks, indicating that fine-tuning makes only modest adjustments to the intermediate representations. This is in sharp contrast to the very low similarity observed between the pretrained and fine-tuned models versus the baseline in these same stages, and reveals that the distinctive message passing pathway of the pretrained model is largely preserved through fine-tuning rather than being restructured toward the baseline solution.

At the Global Decoder, however, the similarity drops substantially, to values of $0.4$ -- $0.9$ depending on the task, indicating that fine-tuning concentrates its representational changes in the final decoding stage. This is consistent with the picture emerging from the second comparison: fine-tuning primarily reorganizes decoder representations to align with the downstream task, while leaving the encoder and message passing representations largely intact. The tasks with the lowest fine-tuned vs pretrained decoder similarity ( $t\bar{t}H(\rightarrow\gamma\gamma)$ CP-even vs CP-odd, FCNC vs $tHq$, and $WH$ vs $ZH$) are the same tasks that show the largest performance improvements from pretraining (Table \ref{tab:results}), suggesting that the decoder reorganization is the key mechanism through which fine-tuning adapts the pretrained model to difficult downstream tasks.

\subsubsection*{Summary}

The three comparisons together reveal a consistent and interpretable picture of how pretraining and fine-tuning interact. The pretrained model begins with encoder representations already closely aligned with the best baseline, which is a strong initialization advantage. It then processes graph information through a fundamentally different message passing pathway that is retained through fine-tuning, suggesting that pretraining instills a distinct and general computational strategy for aggregating physics information. Fine-tuning then primarily reorganizes the final decoder representations to align with the downstream task, leaving the encoder and message passing stages largely unchanged. This picture is consistent with pretraining providing not merely a better parameter initialization, but a richer and more flexible representational structure that fine-tuning can efficiently specialize for diverse downstream tasks.

\subsection{Computational Efficiency}

We evaluate computational efficiency using two ratios of fine-tuned training time to baseline training time, focusing only on the multiclass pretrained model. The first metric is the \textit{time-to-target}, defined as the wall-clock time required to reach a test AUC within $10^{-3}$ of the baseline model's best test AUC. For this metric, we compare the fine-tuned model's time-to-target against the baseline model's time to reach the same target:

\begin{equation}
    \eta_{\mathrm{target}} =
    \frac{T_{\mathrm{target}}^{\mathrm{fine\text{-}tuned}}}
         {T_{\mathrm{target}}^{\mathrm{baseline}}}.
\end{equation}
This represents an optimistic target-aware early-stopping scenario. Of the 135 fine-tuning runs evaluated, only 2 fail to reach the target threshold: one FCNC vs.\ $tHq$ ensemble member at $10^3$ events, which falls short of the target by $0.014$ AUC, and one s-top vs.\ $t\bar{t}H$ ensemble member at $10^3$ events, which falls short by $0.002$ AUC. These two runs are excluded from the time-to-target analysis.

The second metric compares the full training time of the fine-tuned and baseline models:
\begin{equation}
    \eta_{\mathrm{full}} =
    \frac{T_{\mathrm{full}}^{\mathrm{fine\text{-}tuned}}}
         {T_{\mathrm{full}}^{\mathrm{baseline}}}.
\end{equation}
Here, full training time is defined as the wall-clock time required to reach a stopping condition of 5 consecutive epochs with no improvement in training loss.

\begin{figure*}[t]
    \centering
    \includegraphics[width=\textwidth]{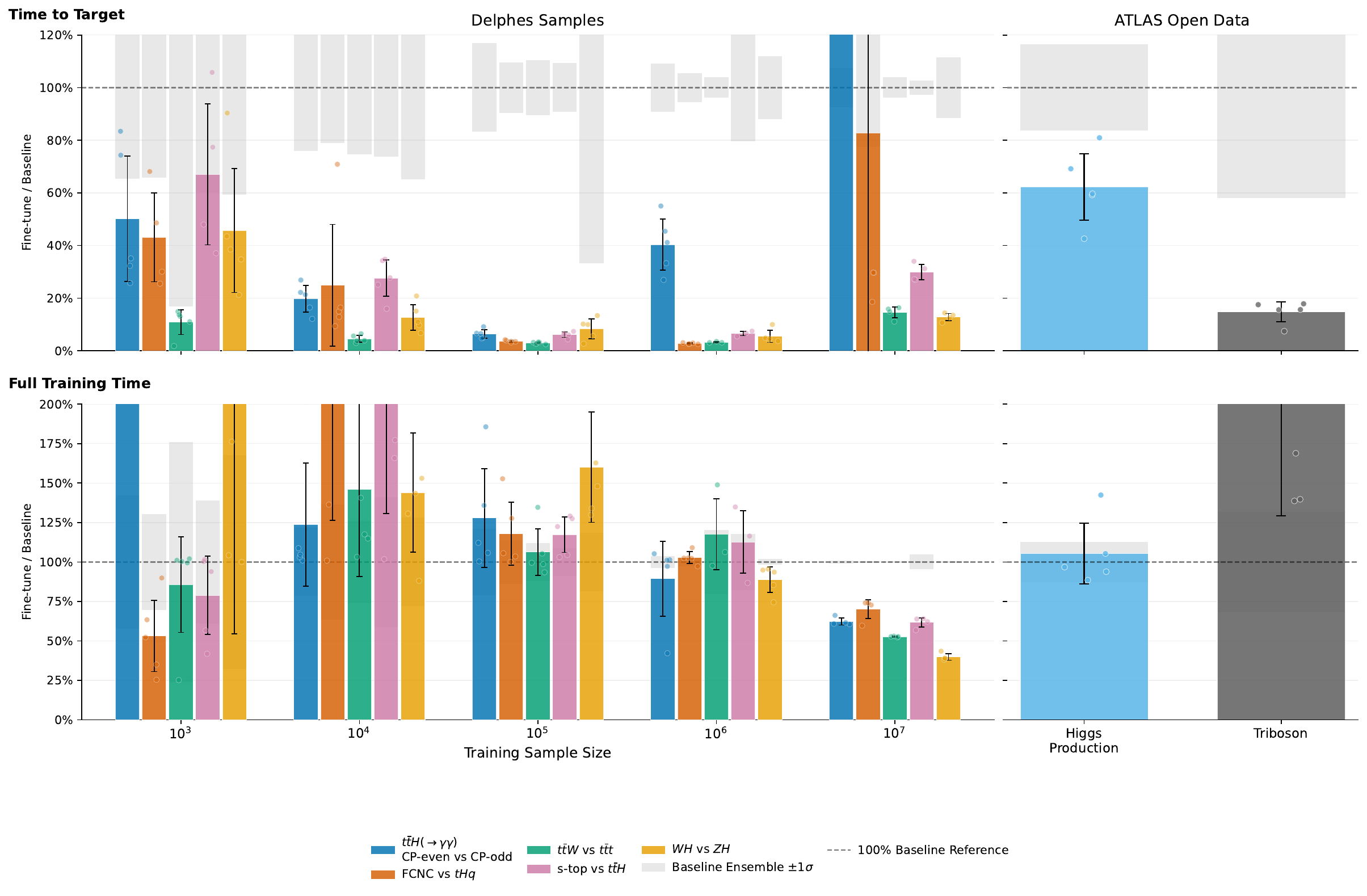}
    \caption{Ratio of multiclass fine-tuned training time to baseline training time, evaluated using two stopping definitions. The top panel uses time-to-target, defined as the wall-clock time required to reach a test AUC within $10^{-3}$ of the baseline model's best test AUC. This represents an optimistic target-aware early-stopping scenario for training. The bottom panel uses full training time, defined as the time required to reach a stopping condition of 5 consecutive epochs with no improvement in training loss. Points show individual ensemble members; bars and error bars show the ensemble mean and standard deviation. Two ensemble members that did not reach the target threshold (one FCNC vs.\ $tHq$ run and one s-top vs.\ $t\bar{t}H$ run, both at $10^3$ events) are excluded from the top panel.}
    \label{fig:time}
\end{figure*}

Figure~\ref{fig:time} shows the two complementary measures of computational efficiency. The top panel reports the time-to-target ratio, which captures the speed at which fine-tuning reaches near-optimal performance and represents an optimistic target-aware early-stopping scenario. The bottom panel reports the full-training-time ratio, which captures the total wall-clock cost of fine-tuning under a standard loss-based stopping condition and represents a more conservative, practical estimate of computational savings.

The time-to-target metric demonstrates the primary advantage of the pretrained model. At $10^5$ events, fine-tuning reaches the AUC target in just $3$--$8\%$ of the baseline training time across all tasks, corresponding to speedups exceeding $12\times$ for every task. This advantage persists across all sample sizes: even for the seven full statistics (5 Delphes and 2 Open Data) tasks, fine-tuning reaches the target in an average of $43.5\%$ of the baseline time. The key characteristic driving this behavior is that the fine-tuned model reaches near-optimal performance rapidly before entering a slow plateau phase, suggesting that aggressive early stopping can reduce computational costs substantially beyond what is reported under the full-training-time metric.

Under the full-training-time metric, fine-tuning can require longer total training time than the baseline at small sample sizes. Because the fine-tuned model uses a learning rate one tenth that of the baseline for all layers except the final linear layer, it converges more slowly and continues training through a plateau long after the AUC target has already been reached, before the loss-based stopping condition eventually triggers. This additional training is not wasted, however: the fine-tuned model typically achieves higher test AUC and accuracy than the baseline by the end of training. This overhead diminishes as the training set grows, and at full statistics the full-training-time ratio falls below unity for every task, with a 7-task mean of $64.6\%$.

The main exception across both metrics is the $t\bar{t}H(\rightarrow\gamma\gamma)$ CP-even vs.\ CP-odd task at $10^7$ events, where the fine-tuned model requires $178\%$ of the baseline time to reach the target, despite ultimately training to completion faster under the full-training-time metric. We attribute this to the inherent difficulty of the task and the absence of photon objects in the pretraining data, consistent with the lower classification performance and larger representational divergence observed for this task throughout the analysis.

Taken together, the two metrics bracket the realistic range of computational savings: the time-to-target ratio gives the best-case saving achievable with well-tuned early stopping, while the full-training-time ratio gives the saving under a standard stopping condition with no additional optimization. Fine-tuning provides consistent time-to-target speedups across all tasks and sample sizes, while the full-training-time advantage is most pronounced at large sample sizes, with fine-tuning being generally slower than the baseline at small sample sizes under the loss-based stopping condition.

To evaluate the total computational cost of the foundation model approach, the pretraining time must be amortized over all fine-tuned tasks. The pretraining times are:

\begin{itemize}
    \item Multiclass pretraining: 45.5 GPU hours
    \item Multilabel pretraining: 60.0 GPU hours
\end{itemize}

The multilabel pretraining time is higher due to a model synchronization step when training in parallel on 16 GPUs. The foundation model approach becomes more computationally efficient than training from scratch once the pretraining cost is offset by the cumulative time savings across fine-tuned tasks.

We compute the crossover point under both efficiency metrics at full statistics ($10^7$ events for Delphes and the full ATLAS Open Data sample) fine-tuned using the multiclass pretrained model. Under the full-training-time metric, baseline training times range from $0.17$ GPU hours (triboson) to $12.24$ GPU hours ($t\bar{t}W$ vs.\ $t\bar{t}t$), with a 7-task mean of $7.77$ GPU hours. Directly averaging the fine-tuned full training time across all tasks gives $4.50$ GPU hours, saving approximately $3.27$ GPU hours per task. Dividing the multiclass pretraining cost of $45.5$ GPU hours by this per-task saving yields a crossover at approximately $14$ tasks.

Under the time-to-target metric, baseline times to target range from $0.19$ GPU hours (triboson) to $2.93$ GPU hours ($t\bar{t}W$ vs.\ $t\bar{t}t$), with a 7-task mean of $1.74$ GPU hours. At a mean time-to-target ratio of $43.5\%$ across all 7 tasks, each fine-tuned task requires on average $0.88$ GPU hours, saving approximately $0.87$ GPU hours per task and yielding a crossover at approximately $52$ tasks. The two crossover estimates of $14$ and $52$ tasks bracket the realistic range depending on how aggressively early stopping is applied.

As a practical example, the ATLAS measurement of Higgs boson couplings using the \hyy\ decay channel~\cite{HIGG-2020-16} involved training 42 classifiers for event categorization, above the full-training estimate of $\sim14$ tasks, though below the time-to-target estimate of $\sim52$ tasks, with the larger time-to-target estimate driven in part by the outlier $t\bar{t}H(\rightarrow\gamma\gamma)$ CP-even vs.\ CP-odd. 
This suggests that the foundation model approach can potentially meaningfully reduce the total computational cost of a realistic high-energy physics analysis, even when pretraining costs are fully accounted for.

\section{Conclusions}
We presented a systematic study of a particle physics foundation model designed to operate on the four-momentum and identification properties of event final-state objects. This model is built on a Graph Neural Network (GNN) architecture and trained on a dataset comprising 120 million simulated proton-proton collision events across 12 distinct physics processes. 

Fine-tuning the pretrained model on downstream classification tasks yields consistent improvements in classification performance relative to models trained from scratch, with the largest gains observed in the low-data regime. The pretrained representations transfer effectively even to processes not seen during pretraining, and performance gains persist at full statistics for most tasks, with the $t\bar{t}H(\rightarrow\gamma\gamma)$ CP-even vs.\ CP-odd task proving the most challenging due to the inherent difficulty of the task and the absence of photon-related events in the pretraining data.
The poor performance of the multilabel pretraining in the low statistics regime further indicates that pretraining objectives must be chosen carefully, as overly prescriptive auxiliary labels can bias the learned representation toward features that are not optimally aligned with downstream event-classification tasks.

The generalizability of the foundation model is further supported by its performance on the ATLAS Open Data tasks. Despite being produced with a different simulation pipeline and detector conditions than the Delphes samples used during pretraining, the model transfers effectively to both tasks, demonstrating that the learned representations are not tied to a specific simulation framework and can generalize across different datasets and event topologies. 
Although the ATLAS Open Data triboson task provides a test involving a different event-generation and detector-simulation setup from the MadGraph+Pythia+Delphes pipeline used for the Delphes samples, we do not perform a controlled study isolating individual generator-level effects, such as alternative parton shower tunes or matrix-element generators for the same physics process. Such controlled generator-variation studies remain an important direction for future work.

The foundation model approach also offers substantial computational advantages. Fine-tuning reaches the AUC target in a small fraction of the baseline training time across nearly all tasks and sample sizes, with the pretraining cost recovered after approximately $14$--$52$ fine-tuned tasks depending on the stopping condition applied. This is well within the scope of a single realistic physics analysis; the ATLAS measurement of Higgs boson couplings in the $H\rightarrow\gamma\gamma$ channel, for example, involved training 42 classifiers for event categorization.

To better understand the learned representations of the pretrained model and guide future optimization efforts, we employed a representational similarity evaluation framework using Centered Kernel Alignment (CKA). This metric allowed us to investigate the source of the performance gains observed in the foundation model.
The layer-wise CKA analysis reveals that fine-tuning preserves the pretrained model's encoder representations while concentrating representational changes in the final decoder stage, suggesting that pretraining provides not merely a better parameter initialization but a richer intermediate computational structure that fine-tuning efficiently specializes for downstream tasks. Future studies could build on this by directly probing whether the distinct message-passing pathways of the fine-tuned and baseline models encode complementary physical information, for example through input feature attribution or ablation studies, which could further inform the design of pretraining objectives and architectures.

\section*{Acknowledgments}
This work is supported by the U.S. National Science Foundation under the Award No.2046280, and by U.S.~Department of Energy, Office of Science under contract DE-AC02-05CH11231.

Joshua Ho would like to thank the support of UC Berkeley Summer Undergraduate Research Fellowships (SURF) and its donors for this work.

We acknowledge the work of the ATLAS Collaboration to record or simulate, reconstruct, and distribute the Open Data used in this paper, and to develop and support the software with which it was analysed.

\bibliography{apssamp}

\appendix

\section{Hyperparameter Optimization}
\label{app:hyperparameter}

\begin{figure*}[t]
    \centering
    \includegraphics[width=\textwidth]{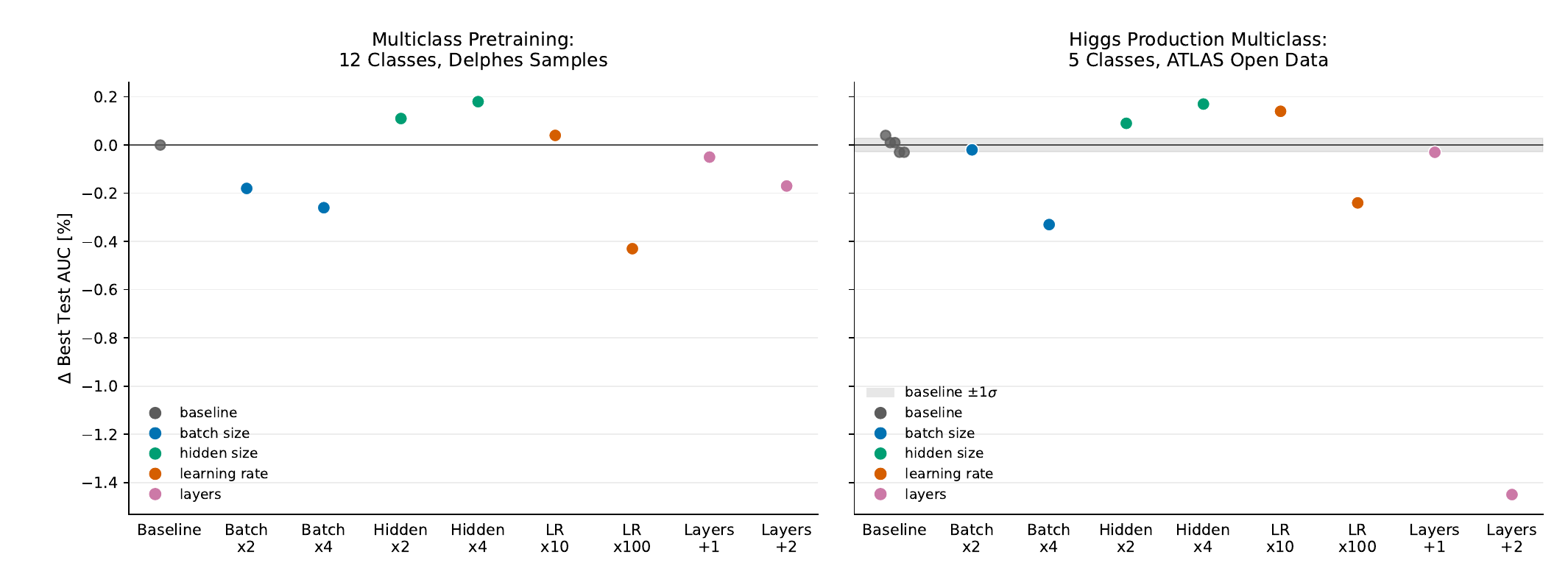}
    \caption{Change in best test AUC relative to the baseline configuration for systematic hyperparameter variations, evaluated on multiclass pretraining with 12 classes on Delphes samples (left) and Higgs production multiclass classification with 5 classes on ATLAS Open Data (right). Each point represents a single training run with one hyperparameter varied at a time. The grey band in the right panel shows the baseline $\pm 1\sigma$ variation across ensemble members. The learning rate $\times 100$ configuration is excluded from the $\pm 0.2\%$ stability claim in the main text as a clear outlier attributable to training instability at excessively large learning rates.}
    \label{fig:hyperparameter}
\end{figure*}

To identify a robust architecture configuration and confirm that the results presented in this work are not sensitive to the choice of hyperparameters, we performed a systematic hyperparameter sweep investing a total of 400 GPU hours on a single NVIDIA A100. The sweep was conducted on two tasks: multiclass pretraining with 12 classes on Delphes samples, and Higgs production multiclass classification with 5 classes on ATLAS Open Data. The latter was chosen as a representative downstream task to ensure that hyperparameter stability holds beyond the pretraining setting.

We swept over four hyperparameter axes independently, varying one parameter at a time relative to the baseline configuration: batch size ($\times 2$, $\times 4$), hidden dimension ($\times 2$, $\times 4$), learning rate ($\times 10$, $\times 100$), and MLP depth ($+1$, $+2$ layers). The results are shown in Figure~\ref{fig:hyperparameter}, where each point shows the change in best test AUC relative to the baseline configuration.

The GNN architecture exhibits broad stability across hyperparameter configurations. For the multiclass pretraining task, all configurations except a learning rate scaling of $\times 100$ remain within $\pm 0.2\%$ of the baseline AUC, with the $\times 100$ learning rate representing a clear outlier at approximately $-0.4\%$, consistent with training instability at excessively large learning rates. The remaining variations show no systematic trend with increasing batch size, hidden dimension, or depth, indicating that the model is not operating near a capacity bottleneck and that the baseline configuration is not a particularly sensitive operating point.

The Higgs production task shows qualitatively similar behavior, with most configurations falling within the baseline $\pm 0.2\%$ band. The $\times 100$ learning rate again performs poorly, and adding two additional layers produces a modest degradation of approximately $-1.4\%$, which may reflect overfitting or optimization difficulty with deeper networks on this task. Configurations with increased hidden dimension and moderate learning rate 
scaling show small positive deviations, suggesting the baseline is slightly conservative in capacity but not meaningfully so.

Based on these results, we adopt the baseline configuration for all experiments reported in this work, as it provides stable and competitive performance without requiring task-specific tuning.

\section{Additional Performance Metrics}
\label{app:performance}

The F1, precision, and recall values are presented for all tasks in Table~\ref{tab:delphes_f1} and Table~\ref{tab:open_data_f1}. These metrics show the same general trends as the accuracy and AUC results described in the body of the paper. For binary outputs, events with scores greater than or equal to $0.5$ are assigned to the positive class, while for multiclass outputs the predicted class is taken to be the class with the largest score. For each class $c$, precision is computed as $P_c = TP_c/(TP_c+FP_c)$, recall as $R_c = TP_c/(TP_c+FN_c)$, and F1 as $F1_c = 2P_cR_c/(P_c+R_c)$, where $TP_c$, $FP_c$, and $FN_c$ are the true positives, false positives, and false negatives for that class. The reported precision, recall, and F1 values are macro-averaged, meaning that the metric is first computed separately for each class and then averaged over classes with equal weight. This gives each class equal importance regardless of its sample size. For each task, pretraining strategy, and training sample size, the metrics are computed separately for each independently trained model and then averaged over all runs.

\begin{table*}[t]
\centering
\begin{tabular}{c c c ccccc}
    \hline
    Name of Task & Metric & Pretraining & \multicolumn{5}{c}{Sample Size} \\
    \cline{4-8}
     & & & $10^3$ & $10^4$ & $10^5$ & $10^6$ & $10^7$ \\
    \hline\hline
    \multirow{9}{*}{$t\bar{t}H(\rightarrow\gamma\gamma)$ CP-even vs CP-odd}
    & \multirow{3}{*}{F1 (\%)}
    & Baseline      & 56.2 $\pm$ 1.0 & 62.2 $\pm$ 0.2 & 64.1 $\pm$ 0.1 & 65.7 $\pm$ 0.0 & 66.2 $\pm$ 0.0 \\
    & & Multiclass   & +3.0 $\pm$ 1.1 & +2.1 $\pm$ 0.2 & +1.1 $\pm$ 0.1 & +0.1 $\pm$ 0.0 & -0.0 $\pm$ 0.0 \\
    & & Multilabel   & +1.3 $\pm$ 1.2 & +1.0 $\pm$ 0.2 & +0.7 $\pm$ 0.2 & +0.0 $\pm$ 0.0 & -0.1 $\pm$ 0.0 \\
    \cline{2-8}
    & \multirow{3}{*}{Precision (\%)}
    & Baseline      & 56.7 $\pm$ 1.1 & 62.5 $\pm$ 0.1 & 64.7 $\pm$ 0.1 & 65.9 $\pm$ 0.0 & 66.4 $\pm$ 0.0 \\
    & & Multiclass   & +2.7 $\pm$ 1.1 & +2.1 $\pm$ 0.1 & +0.7 $\pm$ 0.1 & +0.1 $\pm$ 0.0 & -0.1 $\pm$ 0.0 \\
    & & Multilabel   & +1.0 $\pm$ 1.2 & +0.9 $\pm$ 0.2 & +0.3 $\pm$ 0.1 & -0.1 $\pm$ 0.0 & -0.1 $\pm$ 0.0 \\
    \cline{2-8}
    & \multirow{3}{*}{Recall (\%)}
    & Baseline      & 56.6 $\pm$ 1.0 & 62.4 $\pm$ 0.1 & 64.4 $\pm$ 0.1 & 65.8 $\pm$ 0.0 & 66.3 $\pm$ 0.0 \\
    & & Multiclass   & +2.8 $\pm$ 1.0 & +2.1 $\pm$ 0.1 & +0.9 $\pm$ 0.1 & +0.1 $\pm$ 0.0 & -0.0 $\pm$ 0.0 \\
    & & Multilabel   & +1.1 $\pm$ 1.2 & +0.9 $\pm$ 0.2 & +0.5 $\pm$ 0.1 & -0.0 $\pm$ 0.0 & -0.1 $\pm$ 0.0 \\
    \hline
    \multirow{9}{*}{FCNC vs $tHq$}
    & \multirow{3}{*}{F1 (\%)}
    & Baseline      & 61.4 $\pm$ 0.5 & 65.6 $\pm$ 0.1 & 66.7 $\pm$ 0.3 & 67.9 $\pm$ 0.1 & 66.6 $\pm$ 0.0 \\
    & & Multiclass   & +2.9 $\pm$ 0.8 & +1.7 $\pm$ 0.1 & +1.3 $\pm$ 0.3 & +0.5 $\pm$ 0.1 & -0.1 $\pm$ 0.1 \\
    & & Multilabel   & -2.1 $\pm$ 0.8 & -0.3 $\pm$ 0.2 & +0.7 $\pm$ 0.3 & +0.0 $\pm$ 0.2 & -0.0 $\pm$ 0.1 \\
    \cline{2-8}
    & \multirow{3}{*}{Precision (\%)}
    & Baseline      & 62.2 $\pm$ 0.3 & 65.7 $\pm$ 0.2 & 66.8 $\pm$ 0.1 & 67.8 $\pm$ 0.1 & 66.7 $\pm$ 0.0 \\
    & & Multiclass   & +2.1 $\pm$ 0.7 & +1.5 $\pm$ 0.2 & +1.1 $\pm$ 0.1 & +0.4 $\pm$ 0.1 & -0.0 $\pm$ 0.0 \\
    & & Multilabel   & -2.4 $\pm$ 0.8 & -0.5 $\pm$ 0.2 & +0.5 $\pm$ 0.1 & +0.0 $\pm$ 0.1 & -0.1 $\pm$ 0.0 \\
    \cline{2-8}
    & \multirow{3}{*}{Recall (\%)}
    & Baseline      & 63.1 $\pm$ 0.4 & 65.9 $\pm$ 0.1 & 67.8 $\pm$ 0.0 & 68.9 $\pm$ 0.0 & 67.9 $\pm$ 0.0 \\
    & & Multiclass   & +1.4 $\pm$ 0.7 & +2.4 $\pm$ 0.1 & +1.1 $\pm$ 0.0 & +0.4 $\pm$ 0.0 & -0.0 $\pm$ 0.0 \\
    & & Multilabel   & -2.6 $\pm$ 0.9 & +0.3 $\pm$ 0.1 & +0.3 $\pm$ 0.1 & +0.1 $\pm$ 0.0 & -0.1 $\pm$ 0.0 \\
    \hline
    \multirow{9}{*}{$t\bar{t}W$ vs $t\bar{t}t$}
    & \multirow{3}{*}{F1 (\%)}
    & Baseline      & 75.7 $\pm$ 0.1 & 77.4 $\pm$ 0.1 & 78.9 $\pm$ 0.0 & 79.8 $\pm$ 0.0 & 80.3 $\pm$ 0.0 \\
    & & Multiclass   & +2.7 $\pm$ 0.2 & +2.2 $\pm$ 0.1 & +1.0 $\pm$ 0.0 & +0.4 $\pm$ 0.0 & +0.0 $\pm$ 0.0 \\
    & & Multilabel   & +1.6 $\pm$ 0.1 & +0.5 $\pm$ 0.2 & +0.3 $\pm$ 0.0 & -0.0 $\pm$ 0.1 & -0.1 $\pm$ 0.0 \\
    \cline{2-8}
    & \multirow{3}{*}{Precision (\%)}
    & Baseline      & 75.8 $\pm$ 0.1 & 77.6 $\pm$ 0.0 & 79.0 $\pm$ 0.0 & 79.8 $\pm$ 0.0 & 80.3 $\pm$ 0.0 \\
    & & Multiclass   & +2.6 $\pm$ 0.2 & +2.0 $\pm$ 0.0 & +1.0 $\pm$ 0.0 & +0.3 $\pm$ 0.0 & +0.0 $\pm$ 0.0 \\
    & & Multilabel   & +1.5 $\pm$ 0.1 & +0.6 $\pm$ 0.1 & +0.3 $\pm$ 0.0 & +0.0 $\pm$ 0.0 & -0.1 $\pm$ 0.0 \\
    \cline{2-8}
    & \multirow{3}{*}{Recall (\%)}
    & Baseline      & 75.7 $\pm$ 0.1 & 77.5 $\pm$ 0.1 & 79.0 $\pm$ 0.0 & 79.8 $\pm$ 0.0 & 80.3 $\pm$ 0.0 \\
    & & Multiclass   & +2.7 $\pm$ 0.2 & +2.2 $\pm$ 0.1 & +1.0 $\pm$ 0.0 & +0.4 $\pm$ 0.0 & +0.0 $\pm$ 0.0 \\
    & & Multilabel   & +1.6 $\pm$ 0.1 & +0.6 $\pm$ 0.2 & +0.3 $\pm$ 0.0 & -0.0 $\pm$ 0.1 & -0.1 $\pm$ 0.0 \\
    \hline
    \multirow{9}{*}{s-top vs $t\bar{t}H$}
    & \multirow{3}{*}{F1 (\%)}
    & Baseline      & 82.8 $\pm$ 0.2 & 86.3 $\pm$ 0.1 & 87.5 $\pm$ 0.1 & 88.4 $\pm$ 0.0 & 88.8 $\pm$ 0.0 \\
    & & Multiclass   & +0.4 $\pm$ 0.2 & +1.6 $\pm$ 0.1 & +0.9 $\pm$ 0.1 & +0.3 $\pm$ 0.0 & +0.0 $\pm$ 0.0 \\
    & & Multilabel   & +1.9 $\pm$ 0.2 & +0.8 $\pm$ 0.1 & +0.5 $\pm$ 0.1 & +0.0 $\pm$ 0.0 & -0.0 $\pm$ 0.0 \\
    \cline{2-8}
    & \multirow{3}{*}{Precision (\%)}
    & Baseline      & 83.0 $\pm$ 0.3 & 86.3 $\pm$ 0.1 & 87.6 $\pm$ 0.0 & 88.4 $\pm$ 0.0 & 88.8 $\pm$ 0.0 \\
    & & Multiclass   & +0.3 $\pm$ 0.3 & +1.6 $\pm$ 0.1 & +0.9 $\pm$ 0.0 & +0.3 $\pm$ 0.0 & +0.0 $\pm$ 0.0 \\
    & & Multilabel   & +1.9 $\pm$ 0.3 & +0.8 $\pm$ 0.1 & +0.5 $\pm$ 0.1 & +0.0 $\pm$ 0.0 & -0.0 $\pm$ 0.0 \\
    \cline{2-8}
    & \multirow{3}{*}{Recall (\%)}
    & Baseline      & 82.8 $\pm$ 0.2 & 86.3 $\pm$ 0.1 & 87.5 $\pm$ 0.1 & 88.5 $\pm$ 0.0 & 88.9 $\pm$ 0.0 \\
    & & Multiclass   & +0.4 $\pm$ 0.2 & +1.7 $\pm$ 0.1 & +0.9 $\pm$ 0.1 & +0.3 $\pm$ 0.0 & +0.0 $\pm$ 0.0 \\
    & & Multilabel   & +2.0 $\pm$ 0.2 & +0.8 $\pm$ 0.1 & +0.5 $\pm$ 0.1 & +0.0 $\pm$ 0.0 & -0.0 $\pm$ 0.0 \\
    \hline
    \multirow{9}{*}{$WH$ vs $ZH$}
    & \multirow{3}{*}{F1 (\%)}
    & Baseline      & 50.4 $\pm$ 0.6 & 53.9 $\pm$ 0.2 & 55.4 $\pm$ 0.4 & 57.5 $\pm$ 0.1 & 58.0 $\pm$ 0.0 \\
    & & Multiclass   & +3.5 $\pm$ 0.6 & +2.9 $\pm$ 0.2 & +2.1 $\pm$ 0.4 & +0.4 $\pm$ 0.1 & +0.1 $\pm$ 0.0 \\
    & & Multilabel   & -0.4 $\pm$ 0.7 & -2.5 $\pm$ 1.1 & +0.4 $\pm$ 0.5 & -0.0 $\pm$ 0.1 & -0.1 $\pm$ 0.0 \\
    \cline{2-8}
    & \multirow{3}{*}{Precision (\%)}
    & Baseline      & 51.3 $\pm$ 0.1 & 54.1 $\pm$ 0.1 & 56.1 $\pm$ 0.2 & 57.6 $\pm$ 0.0 & 58.1 $\pm$ 0.0 \\
    & & Multiclass   & +2.7 $\pm$ 0.2 & +2.7 $\pm$ 0.1 & +1.5 $\pm$ 0.2 & +0.3 $\pm$ 0.0 & +0.1 $\pm$ 0.0 \\
    & & Multilabel   & -0.5 $\pm$ 0.2 & +0.3 $\pm$ 0.5 & +0.2 $\pm$ 0.2 & +0.1 $\pm$ 0.1 & -0.0 $\pm$ 0.0 \\
    \cline{2-8}
    & \multirow{3}{*}{Recall (\%)}
    & Baseline      & 51.2 $\pm$ 0.1 & 54.0 $\pm$ 0.1 & 55.8 $\pm$ 0.1 & 57.6 $\pm$ 0.0 & 58.0 $\pm$ 0.0 \\
    & & Multiclass   & +2.8 $\pm$ 0.2 & +2.8 $\pm$ 0.1 & +1.7 $\pm$ 0.1 & +0.4 $\pm$ 0.0 & +0.1 $\pm$ 0.0 \\
    & & Multilabel   & -0.4 $\pm$ 0.2 & -0.5 $\pm$ 0.2 & +0.3 $\pm$ 0.1 & +0.0 $\pm$ 0.0 & -0.1 $\pm$ 0.0 \\
    \hline
\end{tabular}
\caption{F1, precision, and recall of the baseline model versus the change due to fine-tuning from various pretraining tasks. All three metrics are macro-averaged over the classes and are reported as percentages. The Baseline rows give the absolute metric values, while the Multiclass and Multilabel rows give the absolute change in percentage points relative to the baseline. These metrics are averaged over 5 independently trained models with randomly initialized weights and trained on random subsets of the data. One exception is the $10^7$ training, where all models use the same dataset due to limitations on the dataset size. The random subsets are allowed to overlap, but this overlap should be minimal because all models take independent random subsets of $10^7$ events. The results shown here are computed using the held-out test set consisting of one million events per class, with the validation set also composed of one million events per class being used for model checkpoint selection. The errors on the baseline rows are the standard deviations across models, while the errors on the Multiclass and Multilabel rows are propagated errors computed by adding the relevant standard deviations in quadrature.}
\label{tab:delphes_f1}
\end{table*}

\begin{table*}[t]
    \centering
    \begin{tabular}{c c c c c c}
        \hline
        Task & Dataset Size & Pretraining & F1 (\%) & Precision (\%) & Recall (\%) \\
        \hline\hline
        \multirow{3}{*}{Higgs Production}
        & \multirow{3}{*}{2,500,000}
        & Baseline    & 63.0 $\pm$ 0.1 & 62.2 $\pm$ 0.0 & 66.3 $\pm$ 0.0 \\
        & & Multiclass  & +0.1 $\pm$ 0.1 & -0.1 $\pm$ 0.1 & +0.5 $\pm$ 0.1 \\
        & & Multilabel  & -1.4 $\pm$ 0.2 & -1.4 $\pm$ 0.1 & -0.9 $\pm$ 0.1 \\
        \hline
        \multirow{3}{*}{Triboson}
        & \multirow{3}{*}{300,000}
        & Baseline    & 49.7 $\pm$ 3.5 & 55.4 $\pm$ 2.3 & 55.6 $\pm$ 2.8 \\
        & & Multiclass  & +5.6 $\pm$ 3.5 & +2.4 $\pm$ 2.3 & +4.2 $\pm$ 2.8 \\
        & & Multilabel  & -0.3 $\pm$ 3.7 & -3.3 $\pm$ 2.7 & -1.0 $\pm$ 3.5 \\
        \hline
    \end{tabular}
    \caption{F1, precision, and recall for the ATLAS Open Data tasks. All three metrics are macro-averaged over the classes and are reported as percentages. The Baseline rows give the absolute metric values, while the Multiclass and Multilabel rows give the absolute change in percentage points relative to the baseline. Uncertainties are computed across 5 independently trained models. No training sample size scaling study was performed; a single dataset size was used with a 80:10:10 training:validation:test split. The validation set was used for model checkpoint selection while the test set performance is reported in this table.}
    \label{tab:open_data_f1}
\end{table*}

\end{document}